\documentclass[prb,showpacs,twocolumn,aps,superscriptaddress,a4paper]{revtex4-1}
\usepackage{dcolumn,amssymb,amsmath,amsfonts,graphicx,latexsym,color}

\begin{document}

\title{Momentum distribution functions in ensembles: the inequivalence of microcannonical and canonical ensembles in a finite ultracold system}

\author{Pei Wang}
\email{wangpei@zjut.edu.cn}
\affiliation{Institute of Applied Physics, Zhejiang University of Technology, Hangzhou 310023, China}
\date{\today}
\author{Gao Xianlong}
\email{gaoxl@zjnu.edu.cn}
\affiliation{Department of Physics, Zhejiang Normal University, Jinhua 321004, China}
\author{Haibin Li}
\email{hbli@zjut.edu.cn}
\affiliation{Institute of Applied Physics, Zhejiang University of Technology, Hangzhou 310023, China}

\begin{abstract}
It is demonstrated that in many thermodynamic textbooks the equivalence of the different ensembles is achieved in the thermodynamic limit. In this present work we remark the inequivalence of microcannonical and canonical ensembles in a finite ultracold system at low energies. We calculate the microcanonical momentum distribution function (MDF) in a system of identical fermions (bosons). We find that, the microcanonical MDF deviates from the canonical one, which is the Fermi-Dirac (Bose-Einstein) function, in a finite system at low energies where the single-particle density of states and its inverse are finite.
\end{abstract}

\maketitle

\section{Introduction}

The recent progress in ultracold atomic gases experiments provides an important test bed for studying the isolated quantum many-body systems, where these systems are sufficiently weakly coupled to the external environment~\cite{isolated}. The non-equilibrium dynamics after quenching in quantum systems is one of the hot topics. In an isolated system of conserved finite particles, one may ask in what conditions the microcanonical and cannonical ensembles are equivalent?

According to statistical mechanics, the momentum distribution function (MDF) in a canonical ensemble is the Fermi-Dirac (Bose-Einstein) function for a system of identical fermions (bosons)~\cite{chandler}. The Fermi-Dirac (Bose-Einstein) function is first derived in an ideal gas of fermions (bosons), but now they are generally used in interacting systems in terms of Fermi liquid theory~\cite{cross}, where the free fermions (bosons) are replaced by the emergent quasiparticles with renormalized energies. However, the MDF in a microcanonical ensemble is difficult to calculate directly. A textbook derivation of the Fermi-Dirac (Bose-Einstein) function in microcanonical ensembles~\cite{zeghbroeck} is based on an assumption that the MDF is the most probable distribution guaranteed by the principle of the equivalence of ensembles, which states that the macroscopic observables (e.g., the MDF) in canonical and microcanonical ensembles are the same in the absence of long-range interactions in the thermodynamic limit~\cite{huang,touchette03,dauxois,bouchet05}. However, it is obliged to study in which conditions the equivalence of ensembles breaks down in a finite system.

Recently, the MDF in cold-atom systems after a quantum quench causes great interests both in theory and experiments~\cite{greiner,kinoshita,kollar11,eurich11,dora11,manmana07,RigolPRE}. In a system far from integrability, according to the eigenstate thermalization hyperthesis~\cite{deutsch,srednicki,rigol08}, the steady state after a quench can be described by an ensemble of eigenstates~\cite{zhang11,rigol08}, close to each other in energy. In other words, a non-integrable system will thermalize into a microcanonical ensemble~\cite{rigol08}. At the same time, a typical cold-atom system in experiments contains only a few thousand particles~\cite{greiner,hofferberth}, far from the thermodynamic limit. The MDF of microcanonical ensembles in a finite system is now accessible in experiments implemented on ultracold atoms. Then it is necessary to revisit the MDF in microcanonical ensembles when the equivalence of ensembles breaks down in a finite system.

In a finite one-dimensional system, Rigol studied the inequivalence of the canonical and grand-canonical ensembles ~\cite{rigol05}. In a system of non-interacting fermions a numerical comparison between the MDFs in microcanonical and canonical ensembles with equidistant single-particle levels is discussed at some special energies~\cite{schoenhammer}. However, a systematic study of the inequivalence between microcanonical and canonical ensembles is still lack. A particularly interesting case is the low-energy state, when the system energy is comparable with the inverse of the single-particle density of state which is nonzero in a finite system. Obviously, the MDF must be zero when the single-particle energy exceeds the system energy, indicating that it is not a smooth function, different from the Fermi-Dirac (Boso-Einstein) function at finite temperature.

In this paper, we discuss the MDF in microcanonical ensembles. We suppose that the system is finite by setting a finite single-particle density of state. We calculate the microcanonical MDF (MMDF) by relating it to the canonical MDF (CMDF). A well-known relation between canonical and microcanonical ensembles is that canonical ensembles describe the subsystems of a microcanonical ensemble~\cite{goldstein}. In this paper, we show that the MMDF of a system can also be calculated according to its CMDF. We find that the MMDF deviates from the CMDF at low energies, while becoming equivalent at high energies. Our results show the breakdown of the equivalence principle, at the same time provide a way to understand the MDF in cold-atom systems after a quench, which is believed to be the MMDF, while difficult to address directly due to the lack of reliable analytical or numerical methods~\cite{kollath07,manmana07,zhang11,eckstein09,cazalilla06,moeckel}.

The contents of the paper are arranged as follows. In Sect.~\ref{sect:MDF} we introduce the methods to derive the MMDF in a finite system. In Sect.~\ref{sect:MMDF-Fermions} the MMDF for fermions are calculated and the first- and second-order approximations for the MDFs are derived. In Sect.~\ref{sect:MMDF-Bosons} we report the corresponding results for a system of bosons. At last, a concluding section summarizes our results.

\section{Method}
\label{sect:MDF}

Let us suppose a system of identical fermions (bosons) in a canonical ensemble. Its MDF is written as
\begin{equation}\label{fermionbosonfunc}
 n_k = \frac{1}{e^{\beta \epsilon_k}\pm 1},
\end{equation}
where $\beta= 1/k_B T$ is the inverse of the temperature and $\epsilon_k$ the quasiparticle energy. The plus and minus signs ($\pm$) are for fermions and bosons, respectively.

The MDF is the expectation value of the momentum operator $\hat n_{k}$, and can be expressed as
\begin{eqnarray}\label{eq:nk}
 n_k = \sum_{i} \frac{e^{-\beta E_i} n_k(\psi_i)}{Z},
\end{eqnarray}
where $n_k(\psi_i)= \langle \psi_i | \hat n_k |\psi_i \rangle$ and $|\psi_i\rangle$ is the eigenstate with energy $E_i$. The partition function is defined as $Z=\sum_i e^{-\beta E_i}$. The sum is over all the eigenstates. When dividing the range of eigenenergies into small intervals $[E-\frac{\Delta E}{2},E+\frac{\Delta E}{2}]$, the sum over eigenstates in Eq.~(\ref{eq:nk}) is rewritten as
\begin{eqnarray}
 n_k = \sum_{E} \sum_{E_i\in [E-\frac{\Delta E}{2},E+\frac{\Delta E}{2}] }\frac{e^{-\beta E_i} n_k(\psi_i)}{Z},
\end{eqnarray}
where the sum to $E$ is over all the energy intervals. Since $\Delta E$ is taken to be small, the eigenenergies $E_i$ in an interval can be treated as a constant. Thus, the MDF is
\begin{equation}\label{emdfmicro}
 n_k = \sum_{E} \frac{e^{-\beta E}}{Z} \left( \displaystyle \frac{\sum_{E_i\in [E-\frac{\Delta E}{2},E+\frac{\Delta E}{2}] } n_k(\psi_i)}{\Omega(E)} \right),
\end{equation}
where $\Omega(E)$ denotes the number of eigenstates in the interval $[E-\frac{\Delta E}{2},E+\frac{\Delta E}{2}]$ and the partition function becomes $Z=\sum_E e^{-\beta E}$. The term in the bracket of Eq.~(\ref{emdfmicro}) is exactly the average of the microcanonical ensemble of the MDF, denoted as $ \overline{ n_k(E)}_{\textbf{mc}}$.

The definition of microcanonical ensemble demands that the number of eigenstates in the energy interval must be large enough to make the function $ \overline{ n_k(E)}_{\textbf{mc}}$ smooth. This condition can be naturally satisfied by converting the sum to $E$ into an integral over $E$ ranging from $0$ to $\infty$. We set the ground energy to zero and suppose that low energy excitations in the system are gapless. The Eq.~(\ref{emdfmicro}) becomes
\begin{eqnarray}\label{emdfint}
n_k = \frac{1}{Z} \int_0^\infty dE D(E) e^{-\beta E} \overline{ n_k(E)}_{\textbf{mc}},
\end{eqnarray}
where $D(E)$ is the density of eigenstates and the partition function becomes $Z(\beta)=\int_0^\infty dE D(E)e^{-\beta E}$. Converting the summation into an integral smears out the fluctuation of MDFs in the eigenstates. This is equivalent to calculating the average with respect to a microcanonical ensemble. In a system far from integrability, the eigenstate thermalization hyperthesis can be applied~\cite{rigol08}, then the fluctuation of MDFs is absent. In this case, the function $\overline{ n_k(E)}_{\textbf{mc}}$ is equal to the MDF of an eigenstate with energy $E$. The MMDF and the MDF of the eigenstate in systems that the eigenstate thermalization hyperthesis can be applied will not be distinguished next, both denoted by $n_k(E)=\overline{ n_k(E)}_{\textbf{mc}}$.

Comparing Eq.~(\ref{emdfint}) with Eq.~(\ref{fermionbosonfunc}), we finally obtain
\begin{eqnarray}\label{emdfequation}
\int_0^\infty dE D(E) e^{-\beta E} n_k(E) =\frac{Z(\beta)}{e^{\beta \epsilon_k}\pm 1}.
\end{eqnarray}
This equation shows that the Laplace transformation of $D(E)n_k(E)$ is the product of the Fermi-Dirac (Bose-Einstein) function and the partition function determined by $D(E)$. The MDF $n_k(E)$ can then be obtained by an inverse Laplace transformation. The MMDF is a universal function of energy, depending only upon the density of eigenstates in the system.

The equivalence of ensembles can be understood as a result of that the product $D(E) e^{-\beta E}$ should be a $\delta$-like function in the thermodynamic limit. In this paper, however, we consider a finite system, where the single-particle density of states, proportional to the volume of the system, is finite. Then we choose the inverse of the single-particle density of states as the energy unit. We find that the equivalence of ensembles breaks down at low energies by showing that the MMDF is different from the Fermi-Dirac (Bose-Einstein) function.

Next we discuss the MMDF in fermionic and bosonic systems, respectively.

\section{MMDF in a system of fermions}
\label{sect:MMDF-Fermions}

We first discuss a system of identical fermions. The density of eigenstates $D(E)$ is calculated. In the ground state, all the quasiparticle levels lower than the Fermi energy (set to zero) are occupied and the others are empty. The excited eigenstates are classified according to the number of the quasiparticles excited. The total density of eigenstates is expressed as an infinite series,
\begin{equation}
 D(E)= \sum_{n=1}^\infty D_n (E),
\end{equation}
where $D_n(E)$ denotes the density of eigenstates in which $n$ particles are excited. Then $D_n(E)$ is expressed as
\begin{equation}
 D_n(E) = \frac{d}{dE} \int_{\sum_j (\epsilon'_j -\epsilon_j) \leq E } \prod_{j=1}^n d\epsilon_j d\epsilon'_j \mathfrak{d}(\epsilon_j) \mathfrak{d}(\epsilon'_j) \theta(-\epsilon_j)\theta(\epsilon'_j),
\end{equation}
where $\mathfrak{d}(\epsilon_k)$ is the quasiparticle density of states and $\theta(\epsilon)$ denotes the Heaviside function. We concentrate on the low-energy eigenstates, in which only the quasiparticles close to the Fermi level contribute to the function $D_n(E)$. In the wide-band limit~\cite{jauho94}, $\mathfrak{d}(\epsilon_k)$ close to the Fermi level is a constant denoted by $\mathcal{D}$. Then we find $D(E)= \sum_{n=1}^\infty\mathcal{D}\frac{(\mathcal{D}E)^{2n-1}}{(2n-1)!}$.

\begin{figure}
\includegraphics[width=1.0\linewidth]{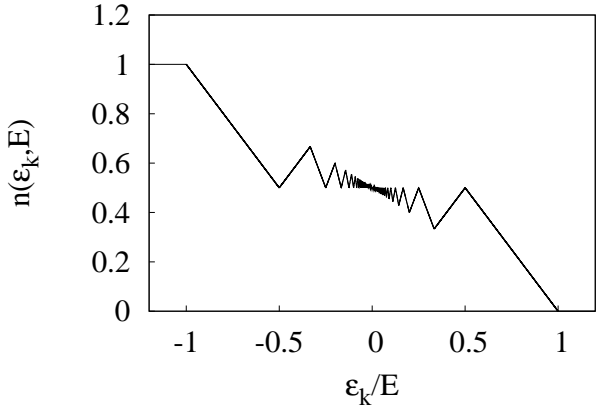}
\caption{The MMDF in the lowest order approximation in a system of fermions. Here $E$ denotes the difference between the system energy and the ground state energy, and $\varepsilon_k$ the quasiparticle energy. The Fermi energy, defined as the energy of the highest occupied quasiparticle state in the ground state, is set to zero.}
\end{figure}
The function $D(E)$ is a power series of $E$. Only the first several terms are important in calculating the MDF at low energies. We first take the lowest order approximation $D(E)= \mathcal{D}^2 E$, and find that the partition function is $Z(\beta)=\frac{{\mathcal{D}}^2}{\beta^2}$ correspondingly. The lowest order approximation is valid in the limit $\beta\to \infty$ or $E\to 0$. By solving Eq.~(\ref{emdfequation}), we find that the corresponding MMDF is a step function:
\begin{eqnarray}\label{fermionemdfzero}
 n(\epsilon_k, E) = \left\{ \begin{array}{cc} j \tilde \epsilon_k, & \mbox{if} \,\,\,\frac{1}{2j+1} \leq \tilde \epsilon_k < \frac{1}{2j} \\ 1- (j+1)\tilde \epsilon_k, & \mbox{if} \,\,\,\frac{1}{2j+2} \leq \tilde \epsilon_k < \frac{1}{2j+1}\end{array}\right.,
\end{eqnarray}
with $j=0,1,2,\cdots$, and rescaled dimensionless energy $\tilde \epsilon_k = \epsilon_k/E >0$ . The MMDF as $\epsilon_k <0$ is given by $n(\epsilon_k,E)=1-n(-\epsilon_k,E)$. In Fig.~1, we plot $n(\epsilon_k,E)$ as a function of $\tilde \epsilon_k$. The MMDF is very different from the CMDF which is a Fermi-Dirac function. It shows quantized features. We find $n(\epsilon_k,E)=1$ for $\epsilon_k<-E$, and $n(\epsilon_k,E)=0$ for $\epsilon_k > E$, respectively. This truncation is natural because the quasiparticle excitation with an energy larger than the eigenstate itself is forbidden. In the regime $-E<\epsilon_k <E$, the MMDF has a sawtooth shape and becomes irregular close to the Fermi energy due to the quantization of quasiparticle excitations in a finite system in which the quasiparticle energies are discrete. In the lowest order, the MMDF depends only upon the ratio $\epsilon_k/E$. It recovers the well-known ground state MDF $\theta(-\epsilon_k)$ in the limit $E\to 0$.

We then consider the second order term in $D(E)$ and have $D(E)=\mathcal{D}^2E+\frac{\mathcal{D}^4 E^3}{6}$. Now the partition function is $Z=\frac{\mathcal{D}^2}{\beta^2} + \frac{\mathcal{D}^4}{\beta^4}$. The MMDF as $\epsilon_k>0$ is found to be
\begin{widetext}
\begin{eqnarray}\label{fermionemdf}
 n(\epsilon_k, E) =\left \{
\begin{array}{cc}
  \left[1-\tilde\epsilon_k (j+1) \right] \displaystyle\frac{1+ \tilde {\mathcal{D}}^2 \left[ 1-(j+1)\tilde\epsilon_k \right]^2/6 + j(j+1)\tilde {\mathcal{D}}^2\tilde \epsilon_k^2/2}{1+\tilde {\mathcal{D}}^2/6}, & \mbox{if} \,\,\,\frac{1}{2j+2} \leq \tilde \epsilon_k < \frac{1}{2j+1} \\ \displaystyle \tilde\epsilon_k j \frac{1+ \tilde {\mathcal{D}}^2\left[ 1-(j+1)\tilde\epsilon_k \right]^2/2 + \tilde {\mathcal{D}}^2\tilde \epsilon_k\left[ 1-(j+1)\tilde\epsilon_k \right]/2 +j^2 \tilde {\mathcal{D}}^2\tilde\epsilon_k^2/6 }{1+\tilde {\mathcal{D}}^2/6}, & \mbox{if} \,\,\,\frac{1}{2j+1} \leq \tilde \epsilon_k < \frac{1}{2j}
\end{array}\right.,
\end{eqnarray}
\end{widetext}
where $\tilde {\mathcal{D}}=\mathcal{D}E$ is a dimensionless quantity. The $n(\epsilon_k,E)$ as $E>0$ depends not only on $\tilde \epsilon_k$ but also on $\tilde {\mathcal{D}}$ after considering the second order term in the density of eigenstates. In Eq.~(\ref{fermionemdf}), the system energy $E$ is contained in $\tilde {\mathcal{D}}$, which increases linearly with $E$. And Eq.~(\ref{fermionemdf}) returns to Eq.~(\ref{fermionemdfzero}) as $\tilde {\mathcal{D}}\to 0$. Since $D(E)$ is a power series of $\tilde {\mathcal{D}}$, the second order approximation is valid as $\tilde {\mathcal{D}}$ is not too large, i.e., $E$ is not much larger than $1/{\mathcal{D}}$. This defines the regime of system energies in which Eq.~(\ref{fermionemdf}) is correct.
\begin{figure}
\includegraphics[width=1.0\linewidth]{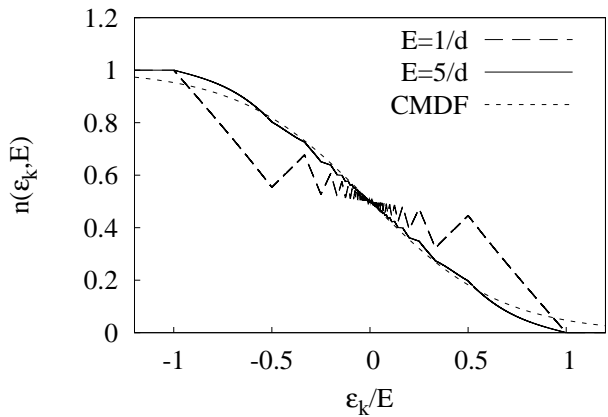}
\caption{The MMDF at different energies in a system of fermions. Here $E$ denotes the difference between the system energy and the ground state energy, and $\varepsilon_k$ the quasiparticle energy. And $\varepsilon_k=0$ denotes the Fermi energy. The canonical MDF at the effective temperature $T=E/(3k_B)$ is plotted as a comparison, labeled by the acronym CMDF.}
\end{figure}
In Fig.~2, we plot the MMDF in Eq.~(\ref{fermionemdf}) at different $E$. As the energy increasing, the sawtooth shape in the MDF is softened. The MMDF gradually changes into the Fermi-Dirac function, i.e., the CMDF. In fact, the difference between the MMDF and CMDF at high energies will totally disappear after considering higher order terms in $D(E)$, a result of the equivalence of ensembles. We take the large $E$ limit ($\tilde {\mathcal{D}}\to \infty$) in Eq.~(\ref{fermionemdf}), and then take the limit $\tilde \epsilon_k \to 0$, and find $n(\epsilon_k,E)=\frac{1}{2}-\frac{3}{4} \tilde \epsilon_k + O(\tilde \epsilon_k^3)$. The slope of the MMDF at the Fermi energy is $-3/(4E)$. We define an effective temperature so that the slope of the corresponding Fermi-Dirac function at the Fermi energy is $-3/(4E)$. The effective temperature is found to be $T = E/(3k_B)$. Because Eq.~(\ref{fermionemdf}) is valid at low energies, this effective temperature should be attached to microcanonical ensembles of a system of fermions at the low-energy limit.

The MDF in microcanonical and canonical ensembles are very different at low temperatures. We have expressed $D(E)$ in a power series of $E$. Correspondingly, $n(\epsilon_k,E)$ is a power series of $\tilde {\mathcal{D}}$. The result in Eq.~(\ref{fermionemdf}) is kept to second order of $\tilde {\mathcal{D}}$. As $\tilde {\mathcal{D}}$ is small, the $O(\tilde {\mathcal{D}}^2)$ terms in Eq.~(\ref{fermionemdf}) is negligible. The MMDF has the sawtooth shape similar to that in Eq.~(\ref{fermionemdfzero}), clearly distinguished from the Fermi-Dirac function. At low energies, the quantization effect of quasiparticle excitations is important, resulting the oscillation behavior of the MDF in the regime $-E < \epsilon_k <E$.

The steady MDF of a nonintegrable system driven out of equilibrium is believed to be the MMDF~\cite{rigol08}. And the corresponding energy $E$ is decided by the initial condition. While, the CMDF is obtained in an open system contacting with a large thermal reservoir. In this paper, we find that the equivalence of ensembles is broken in a finite system at low energies $E\sim 1/{\mathcal{D}}$. The inequivalence of ensembles has been intensively discussed in classical mechanics in the presence of long-range interactions~\cite{touchette03,dauxois,bouchet05,murata,cohen}. Our results must be distinguished from those discussions, because the inequivalence of ensembles in our paper is a finite-size effect due to the quantization of excitations. The equivalence will be recovered at high temperatures.

Our results are related to experiments on isolated ultracold atoms, in which the MMDF can be directly measured. The MMDF in Eq.~(\ref{fermionemdf}) can be observed in a system in which the CMDF is the Fermi-Dirac function, e.g., in a Fermi gas or a Fermi liquid. This suggests that the system is two- or three-dimensional, since the one-dimensional fermions are described by a Luttinger liquid~\cite{delft98}. Here we discuss the condition that the inequivalence of ensembles is obvious in two dimensions. The Fermi gas in two dimensions has been realized in experiments (see Refs.~\cite{orel,feld,martiyanov} for examples). We choose an isotropic dispersion relation $\epsilon_k = \frac{\hbar^2 k^2}{2m}$, where $m$ denotes the quasiparticle mass. In two dimensions, the quasiparticle density of states is a constant as ${\mathcal{D}}=\frac{mS}{2\pi \hbar^2}$, where $S$ is the area of the system. Then we have $\tilde {\mathcal{D}}=\frac{E mS}{2\pi \hbar^2}$. We use the relation between the temperature and the system energy obtained above and finally we get $T=\frac{2\pi\hbar^2 \tilde {\mathcal{D}}}{3k_B mS}$. The inequivalence of ensembles is obvious as $\tilde d\sim 1$. Then the temperature must be inversely proportional to the system area.

When considering a system of $^{40}$K atoms~\cite{feld} with the mass $6.6\times 10^{-26} kg$ and a typical area of two-dimensional Fermi gas $S=1\mu m^2$, we obtain the typical temperature $T \sim 2.5\times 10^{-8}K$ where the inequivalence of ensembles can be observed.

\section{MMDF in a system of bosons}
\label{sect:MMDF-Bosons}
Next we discuss a system of bosons. Without the Pauli exclusion principle, the infinite number of bosons occupies the lowest quasiparticle level (set to zero) in the ground state. Then the density of excited eigenstates becomes
\begin{equation}
D(E) = \sum_{n=1}^\infty \frac{d}{dE} \int_{\sum_j \epsilon_j \leq E } \prod_{j=1}^n d\epsilon_j \mathfrak{d}(\epsilon_j) .
\end{equation}
Here we consider the case that the quasiparticle density of states is a constant ${\mathcal{D}}$, e.g., in two-dimensional systems. Then the density of eigenstates is found to be $D(E)=\sum_{n=1}^\infty \frac{{\mathcal{D}}^n E^{n-1}}{(n-1)!}$. Again, the function $n(\epsilon_k,E)$ is a power series of $\tilde {\mathcal{D}}={\mathcal{D}}E$ and can be calculated order by order at low energies. We keep $D(E)$ to order $O(E^2)$: $D(E)={\mathcal{D}}+{\mathcal{D}}^2 E + {\mathcal{D}}^3 E^2/2$. This corresponds to a function $n(\epsilon_k,E)$ kept to order $\tilde {\mathcal{D}}^2$. The partition function is $Z=\frac{{\mathcal{D}}}{\beta}+\frac{{\mathcal{D}}^2}{\beta^2} + \frac{{\mathcal{D}}^3}{\beta^3}$. The Laplace transformation in the case of bosons is defined as
\begin{equation}
\int^\infty_0 dE e^{-\beta E} D(E) n(\epsilon_k, E) = \frac{Z(\beta)}{e^{\beta \epsilon_k}-1}.
\end{equation}
Performing the inverse Laplace transformation, we find
\begin{widetext}
\begin{equation}
 n(\epsilon_k,E)= j \frac{1+\tilde {\mathcal{D}} - \frac{j+1}{2} \tilde {\mathcal{D}} \tilde \epsilon_k +\frac{\tilde {\mathcal{D}}^2}{2}\left( (1 - j\tilde \epsilon_k)^2 + (j-1) \tilde \epsilon_k (1-j\tilde \epsilon_k) + \frac{(j-1)(2j-1)}{6} \tilde \epsilon_k^2 \right) }{1+\tilde {\mathcal{D}} +\tilde {\mathcal{D}}^2/2},
\end{equation}
\end{widetext}
as $\frac{1}{j+1} \leq \tilde \epsilon_k=\epsilon_k /E < \frac{1}{j}$ with $j=0,1,2,\cdots$.

\begin{figure}
\includegraphics[width=1.0\linewidth]{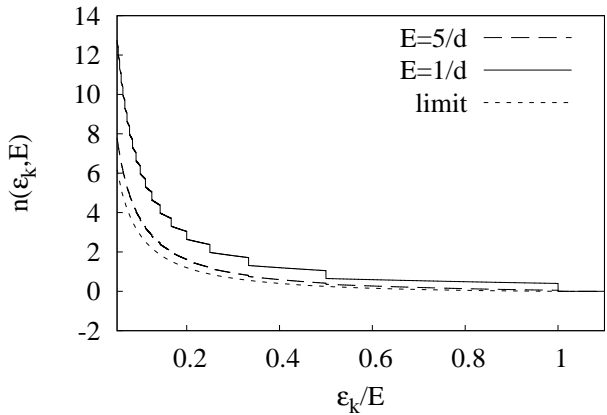}
\caption{The MMDF in a system of bosons at different energies. Here $E$ denotes the difference between the system energy and the ground state energy, and $\varepsilon_k$ the quasiparticle energy. And $\varepsilon_k=0$ denotes the energy of the lowest quasiparticle state. The function $n(\epsilon_k,E)=\frac{1}{3\tilde \epsilon_k} -\frac{1}{2} + \frac{\tilde \epsilon_k}{6}$ in the limit $\tilde {\mathcal{D}}\to \infty$ is plotted as a comparison, labeled by 'limit'.}
\end{figure}
In Fig.~3, we plot the bosonic MMDF at different energies. The MMDF is a step function at low energies, while changes into a smooth one as the energy increasing. Taking the limit $\tilde {\mathcal{D}}\to \infty$, we get $n(\epsilon_k,E)=\frac{1}{3\tilde \epsilon_k} -\frac{1}{2} + \frac{\tilde \epsilon_k}{6}$. This indicates that at low energies the bosonic MMDF as a function of quasiparticle energy decreases in a power way instead of in an exponential way suggested by the Bose-Einstein function. The condition of observing this power law decay is that $\tilde {\mathcal{D}}$ is not too large so that the second order approximation to $n(\epsilon_k,E)$ is effective.

\section{Conclusions}
\label{sect:Conclusions}
In summary, we calculate MMDFs in a system of fermions and bosons. The MMDF given in this paper is universal in systems where the corresponding CMDF of quasiparticles is the Fermi-Dirac (Bose-Einstein) function and the low energy excitations are gapless. The MMDF is found to be different from the CMDF at low energies as $E\sim 1/{\mathcal{D}}$, where ${\mathcal{D}}$ is the quasiparticle density of states. It shows a sawtooth shape for fermions and a power-law decay for bosons. Our results show the breakdown of the equivalence of ensembles, as a result of the quantization of quasiparticle excitations which can only be observed in small systems, since ${\mathcal{D}}$ is proportional to the system volume. The MMDF obtained in this paper is expected to be verified in experiments of isolated cold-atom systems, which is believed to thermalize into a microcanonical state after a quench.

\section{Acknowledgements}
\label{acknowledgements}
Gao X. was supported by the NSF of China under Grants No. 11174253 and by the Zhejiang Provincial Natural Science Foundation under Grant No. R6110175.

\end{document}